\documentstyle[epsfig,12pt]{article}

\if@twoside
\else
\fi
\advance\textheight by \topskip
\def\swein{\sin^2\theta_W^{\rm eff}}
\def\alr{A_{LR}}
\def\ee{e^+e^-}
\def\pole{{\cal P}_e}

\def\pola{\langle\!{\cal P}_e\!\rangle}

\def\alum{A_{\cal L}}
\def\aeff{A_\varepsilon}
\def\apol{A_{\cal P}}

\def\zbb{$Z$-$b$-$\bar{b}$}

\begin{document}
\rightline{OREXP 97-03}
\bigskip
\begin{center}
{\LARGE Recent Electroweak Results from SLC/SLD\footnote{
\it Presented at Beyond the Standard Model V, Balholm, Norway, 
April 29 -- May 4, 1997.} } \\[0.2in]
{\large Raymond E. Frey}\\[0.05in]
{\large\sl representing the SLD Collaboration} \\[0.15in]
{\it Physics Department, University of Oregon\footnote{Supported by DoE
award DE-FG03-96ER40969.} \\
Eugene, Oregon 97403} \\
{E-mail: rayfrey@bovine.uoregon.edu}
\end{center}

\begin{abstract} \noindent
Key electroweak measurements performed by the
SLD collaboration at the SLC are described and recent results given.
The left-right cross-section asymmetry, $A_{LR}$, has been updated
to include the 1996 data. It remains the single most precise measurement
of $\swein$ and it is compared to the LEP results. The polarized differential
cross section for $b$-quarks is measured and is used to perform a
unique direct measurement of the parity violation parameter for $b$ quarks,
$A_b$. The excellent capability to perform secondary vertexing at
SLC with CCD-based vertex detectors is described, including first
physics results with the new detector VXD3. The vertexing is
used to full advantage to make a highly pure $B$ tag to measure the fraction
of hadronic $Z^0$ decays going to $b$ quarks, $R_b$. The vertexing, in
combination with electron-beam polarization, is used to measure $B_d^0$ mixing.
The prospects for making a $B_s^0$ mixing measurement are excellent
given good SLC performance in the upcoming SLC run(s).
\end{abstract}

\section{Introduction}

Two key elements of $Z^0$ physics measurement at the SLAC Linear Collider
(SLC) have made possible a set of unique precision electroweak tests 
of the Standard Model (SM).  Highly longitudinally polarized electron beams
directly probe the parity-violating nature of electroweak physics,
greatly enhancing the measurement of key parameters
such as $A_e$ and $A_b$. In addition, the inherent interaction point properties
of linear $e^+e^-$ colliders, namely a tiny and very stable interaction region
and a low collision rate,  have allowed the SLD experiment to push the
technique of precision vertexing to a new level. 
These assets have enabled SLD to make a set of SM tests at
comparable precision to those of LEP, while using largely independent
measurement techniques.

\subsubsection*{Polarized Beam}
\label{sec:slc-pol}

\vskip -0.1in
The production and transport of longitudinally polarized electron beams
at SLC has been discussed in detail elsewhere\cite{ref:pol}.
Polarized electrons are photoproduced from a strained
GaAs cathode by means of circularly polarized laser light with
$\lambda\approx 850$ nm. The sign of the laser polarization is controlled
such that it follows a pseudo-random pulse-by-pulse sequence.
Before the electron beam enters its damping ring, the polarization is rotated
into the vertical by means of a $6.34$ T-m solenoidal field. 
As it turns out, the wavelength of betatron oscillations of the collider arc 
is very nearly equal to the distance for electron spin-precession. Hence,
spin precession in a non-bend plane can accumulate during the normal arc
focussing-defocussing beam transport. This allows the vertical polarization
at the end of the linac to become longitudinal at the collision point by
means of the beam ``launch'' in position and angle at the entrance to the arc. 
The launch
can be carefully tuned at the same time that polarization measurements
at SLD are made.  A thorough optimization is carried out in this way
during initial running setup, and is subsequently checked and re-optimized
periodically during the run. Occasionally, changes of beam 
orbit in the arc due to normal accelerator tuning lead to
small changes in polarization ($\sim 0.5\%$) with timescales of hours.
Meanwhile, polarization measurements
are performed every 3 minutes during normal data taking, with each
measurement having statistical resolution of $\approx 0.5\%$.
Hence, any modulations in polarization are recorded and applied 
to the corresponding (in time) $Z^0$ events.
A summary of the polarization measurement and systematic effects 
is given below in the $\alr$ section.

The electron and positron beams at SLC are each focussed at the collision
point to a typical size of $(\sigma_x,\sigma_y,\sigma_z)=(2.1,0.6,500) \mu$m.
This allows for a small-diameter beampipe and vertex detectors which
can be placed close to the collision point, thus reducing extrapolation error.
Equally important is the fact that the collision point is very stable. Its
transverse movement is usually less than a few microns between measurements
of the position, which requires a few tens of $Z^0$ events. The resulting accuracy
for determining the interaction point position is $\approx 7$ $\mu$m
in $x$ and $y$, and $\approx 40$ $\mu$m in $z$. Not only
does this significantly improve the precision for measuring decay lengths or
impact parameters, it also means that particles which decay after short flight
paths can be separated from the primary interaction point. Besides improving
the efficiency for tagging $b$-quark and $c$-quark flavored hadrons and $\tau$
leptons, the $Z^0$ decay modes involving these final states can be vetoed to
allow studies of enhanced samples of $Z^0$ decays to $u$ and $d$ quarks. The
latter principle of light-flavor tagging has used at SLD for QCD 
measurements, but is not discussed further here.
The low 120 Hz collision rate at SLC opens a window of opportunity for
exploiting an intrinsic 3-D pixel technology for the vertex detector. In particular, 
CCD-based detectors have been developed and utilized for SLD, starting with
VXD2 in 1992. 

\subsubsection*{VXD3: The Vertex Detector Upgrade}
\label{sec:vxd3}

\begin{figure}[tbh]
\begin{center}
\makebox[2.9in]{
\epsfxsize=3.1in
\epsfbox{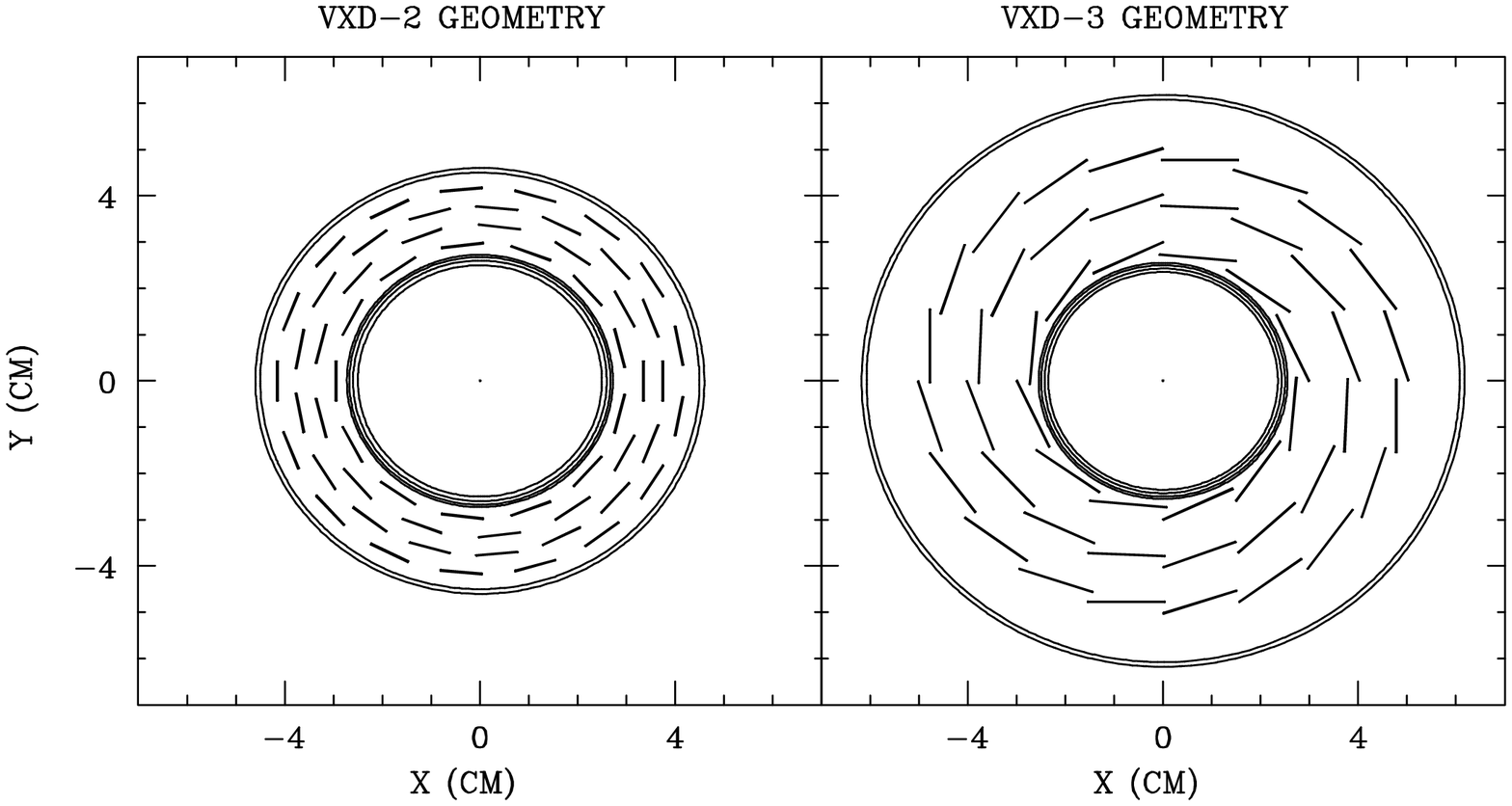}
 } \makebox[2.9in]{
\epsfxsize=2.9in 
\epsfbox{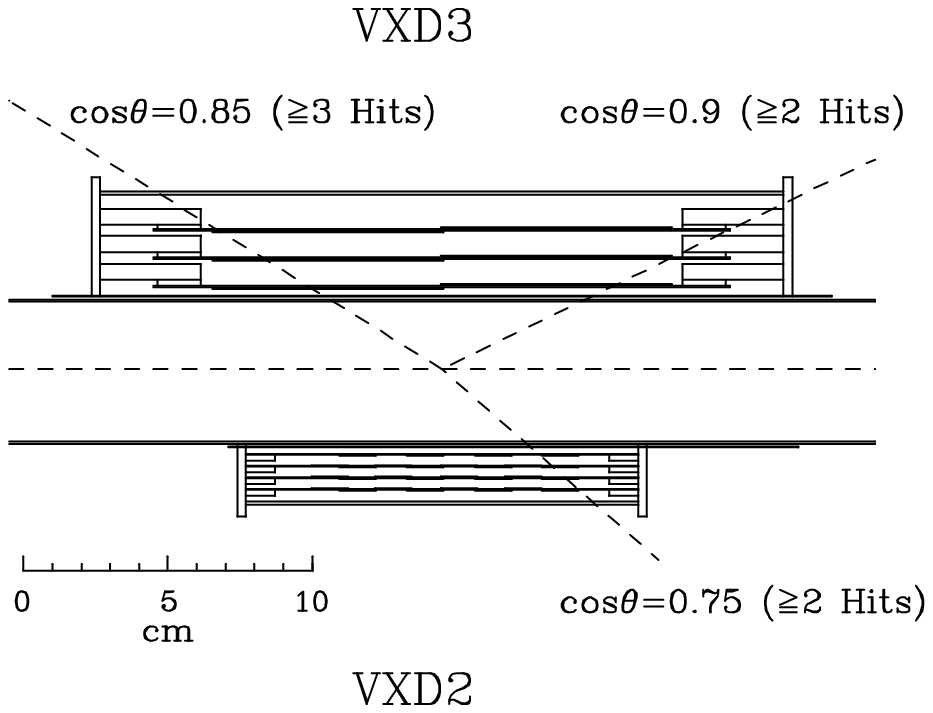} }
\end{center}
  \caption{Geometry of VXD3 upgrade, $x$-$y$ view (left) and
    $r$-$z$ view (right).} 
  \label{fig:VXD3-geom}
\end{figure}

\vskip -0.1in
Full realization of the physics potential of 
the CCD-based VXD2 vertex detector was achieved after a few
years of experience.
It soon became clear that an improved detector, exploiting the rapid
improvement in commercially available CCDs, could be made. The most
obvious characteristic of the upgrade to VXD3 can be seen in Fig. 
\ref{fig:VXD3-geom}, namely the expanded coverage both transversly
and longitudinally. This was made feasible because it had become
possible to fabricate individual CCDs of much larger size.
The transverse geometry was improved in two major aspects:
The layout of the CCD ladders was configured such that
all tracks would have three good hits, one in each of the three layers. Secondly,
the expanded distance between layers provides an improved 
resolution for track extrapolation
inward to the collision point and outward to match the tracks found
in the central drift chamber.  From the righthand diagram of 
Fig.~\ref{fig:VXD3-geom}
it is clear that VXD3 also extends the coverage in $\cos\theta$ significantly,
with the 3-hit and 2-hit vertexing capability extending to $|\cos\theta |$
of $0.85$ and $0.90$, respectively. This is a substantial improvement relative
to VXD2 and has important physics implications, particularly with regard to
the polarized forward-backward $b$-quark asymmetry. The large 
forward-backward asymmetry for $b$ {\it vs} $\bar{b}$ with polarized
beam means that the forward coverage is very important not only for the
measurement of $A_b$, but also as an initial-state tag for $B^0$ mixing
measurements. This is discussed later.

\vskip -0.25in

\begin{table}[tbh]
\begin{center}
\parbox{5in}{\caption{Comparison of VXD2 and VXD3 parameters. 
$\sigma_b$ is impact parameter resolution with $p$ in GeV/c.}\smallskip}
\begin{tabular}{|c|c|c|} \hline
Parameter & VXD2 (1992--95)& VXD3 (1996-- ?) \\ \hline
\# CCDs & 480 & 96 \\
\# pixels & $1.1\times 10^8$ & $3.1\times 10^8$ \\
Rad. Lengths & $4.1\%$ & $2.1\%$ \\
$\sigma_b(xy)$ ($\mu$m) & $11\oplus 70/p$ & $9\oplus 29/p$ \\
$\sigma_b(rz)$ ($\mu$m) & $38\oplus 70/p$ & $14\oplus 29/p$ \\ \hline
\end{tabular}
\end{center}
\label{tab:vxd}\end{table}\vskip -0.3in

Table \ref{tab:vxd}~summarizes a few key parameters of the
upgrade of VXD2 to VXD3. Most of the SLD data has been recorded
with VXD2. However, we present here preliminary results for the 1996 run
with VXD3. The improvements in coverage and resolution have a clear
impact, as the $b$ and $c$-quark measurement errors for the 1996 data,
while representing only about $25\%$ of the data,
are in many cases comparable to those of the VXD2 period.

\medskip

SLD also has a powerful particle identification system which has made
a noticeable impact on the electroweak physics. This Cherenkov Ring
Imaging Detector (CRID) is similar in concept to the system used by
DELPHI at LEP, and the results are of similar quality. Its use for electroweak
physics has been to tag $K^{\pm}$, and hence identify $b$ {\it vs} $\bar{b}$
via the $b\rightarrow c\rightarrow s$ cascade. Charged particle momentum measurement
is made by the central drift chamber which is emersed in a uniform $0.6$ T
solenoidal field. Electron identification and total energy measurement is
performed by a finely-segmented liquid-argon calorimeter (LAC) which provides
complete coverage for $|\cos\theta |<0.98$.
Muon identification and magnetic field
return takes place in the ``warm'' iron calorimeter (WIC).

\section{The Left-Right Cross-Section Asymmetry}
\label{sec:alr}

The left-right cross-section asymmetry is defined as
\begin{equation}
\alr\equiv\left(\sigma_L-\sigma_R\right)/
\left(\sigma_L+\sigma_R\right), 
\end{equation}
where $\sigma_L$ and $\sigma_R$ are the $\ee$ production
cross sections for $Z$ bosons
with left-handed and right-handed electrons, respectively.  
To leading order, the Standard Model predicts
that this quantity depends upon the vector ($v_e$)
and axial-vector ($a_e$) couplings of the $Z$ boson to the electron
current,
\begin{equation}
\alr=A_e={2v_ea_e\over v_e^2+a_e^2}
={2\left[1-4\swein\right]\over1+\left[1-4\swein\right]^2}
\label{eq:alr}\end{equation}
We retain the tree-level relations between the couplings and
the electroweak mixing parameter to compare with the experimental
measurement of $\alr$, effectively including higher-order
corrections in $\swein$ (as well as $a_e$ and $v_e$). This
effective electroweak mixing parameter is then defined by
\begin{equation} 
\swein\equiv(1-v_e/a_e)/4 
\end{equation}
This convention is particularly
useful for the measurement of the electroweak asymmetries at
the $Z$ resonance. 

The $\alr$ event selection is very simple and is designed to
efficiently select a pure sample of hadronic $Z$ decays. 
Tau-pair and muon-pair final states
are not explicitly excluded in the selection, but the efficiency is small. 
Bhabha events ($e^+e^-$ final states) are excluded due to the complications involved
in identifying and interpreting the large $t$-channel contribution. 
In the LAC we require $ E_{tot}\equiv\sum E_i > 15$ GeV and
$ E_{imb}\equiv\vert\sum E_i \hat{r}_i \vert / E_{tot} < 0.6$,
where the sums are over LAC clusters $i$ having energy greater
than 100 MeV and $\hat{r}_i$ is a unit displacement
vector from the collision point to the $i$th LAC cluster.
The only other criteria requires a few good tracks in the drift chamber:
at least 3 tracks in one hemisphere or at least 2 tracks in both hemispheres. 
These criteria eliminate Bhabha, radiative Bhabha, 2-photon, and beam
background events, resulting in a sample purity of $99.89\pm 0.08\%$ with
a selection efficiency of $89.3\pm 0.8\%$.

The $\alr$ error is statistics dominated. By far the leading systematic
error is that of the polarization measurement. Table \ref{tab:pol-sys}
summarizes the polarization measurement errors for each SLC/SLD run.
The entry labelled ``AP'' refers to the polarimeter analyzing power, that
is, the expected Compton cross-section asymmetry for a fully-polarized beam.
The kinematic cutoff for Compton scattering corresponds to electrons fully
backscattered in the c.o.m. frame, which is also where the asymmetry is
maximum. This provides a natural means for calibrating the
polarimeter to the physical scale.
The entry labelled ``chromaticity'' refers to a small effect which, due
to imperfect beam chromatic corrections, allow the  
polarization at the $e^+e^-$ IP to be slightly
different than that measured at the $e^-\gamma$ IP some 25m away. Before the
1994-95 run a number of steps were taken to make this effect small,
including changes to the electron-beam bunch compression and better and
more frequent measurements of beam dispersion.
For the 1996 run a second polarimeter (PGC) was installed to crosscheck
the Compton polarization measurement at the $\sim 1\%$ level
by measuring the Compton photon
energy asymmetry, rather than the Compton electron asymmetry as with
the standard polarimeter. Preliminary results show that the measurements
agree to better than $1\%$.

\vskip -0.25in
\begin{table}[htbp]
\begin{center}
\parbox{5.0in}{\caption{Systematic uncertainties on the polarization measurement,
    $\delta\pole/\pole$. The uncertainty for the 1996 run given below is inflated
    due to the preliminary nature of the analysis at the time of presentation. Subsequent
    analysis indicates a final error which again will be $\approx 0.7\%$.
    } \smallskip}
\begin{tabular}{|l||c|c|c|c|} \hline
 & 1992 & 1993 & 1994-95 & 1996 \\ \hline
Laser Polarization & $2.0\%$ & $1.0\%$ & $0.20\%$ & $0.20\%$ \\
Detector Linearity & $1.5\%$ & $1.0\%$ & $0.50\%$ & $0.50\%$ \\
AP Calibration       & $0.4\%$ & $0.5\%$ & $0.29\%$ & $0.30\%$ \\
Electronic Noise    & $0.4\%$ & $0.2\%$ & $0.20\%$ & $0.20\%$ \\
Inter-channel consistency & $0.9\%$ & $0.5\%$ & --- & $0.8\%$ \\ \hline
Total Polarimeter Uncertainty & $2.7\%$ & $1.6\%$ & $0.67\%$ & $1.03\%$ \\
Chromaticity ($\xi$) & --- & $1.1\%$ & $0.17\%$ & $0.18\%$ \\ \hline
Total $\pole$ Uncertainty & $2.7\%$ & $1.9\%$ & $0.69\%$ & $1.04\%$ \\ \hline
\end{tabular} \end{center}
\label{tab:pol-sys}
\end{table}

\vspace{-0.3in}
\subsection{Results}

A complete expression which summarizes the $\alr$ measurement is
\begin{eqnarray}
\alr & = &{A_m\over \pola} + {1\over\pola}\{ f_b(A_m-A_{\rm bkg}) - \alum
+ A_m^2\apol \nonumber \\
& - & E{\sigma^\prime(E) \over\sigma(E)} A_E - \aeff + \pola{\cal P}_p\}
\end{eqnarray}\label{eq:alr-corr}
The quantity $A_m$ is the measured asymmetry in terms of numbers of events:
$A_m = (N_L-N_R)/(N_L+N_R)$. 
The terms within the curly brackets represent very small, nearly negligible,
corrections. The larger of these terms, the background correction ($f_b$), 
luminosity asymmetry ($\alum$), and polarization asymmetry ($\apol$), 
are summarized in Table \ref{tab:alr-corr}.
Note that all quantities are given in units of $10^{-4}$. 
The individual polarization measurements 
${\cal P}_i$ associated with each $Z^0$ event are combined to form the
luminosity-weighted average polarization:
\vskip -0.25in
\begin{equation} 
\pola = (1+\xi) {1\over N_Z}\sum_i^{N_Z} {\cal P}_i \,,
\end{equation} 
where $\xi$ is the chromatic correction discussed above.
Table \ref{tab:alr-summary} summarizes the $\alr$ 
measurements\cite{ALR-PRL93,ALR-PRL94,ALR-PRL97} for each
SLC/SLD run, and a complete description of all terms can be found in the references.

\begin{table}[htbp]
\begin{center}
\parbox{5in}{\caption{Terms from Eq. 4 in units of $10^{-4}$. The terms not
shown are considered negligible. The corrections actually applied to $A_m$ are
given in the bottom row. The 1996 results are preliminary.}\smallskip}
\begin{tabular}{|l||c|c|c|c|} \hline
Quantity & 1992 & 1994 & 1994-95 & 1996 \\ \hline
$A_m$    &               $223\pm 99$ & $1031\pm 46$ & $1144\pm 32$ & $1178\pm 44$ \\ \hline
$f_b(A_m-A_{\rm bkg})$ & $4.7\pm 1.6$ & $1.75\pm 0.72$ & $0.65\pm 0.55$ & $0.65\pm 0.55$ \\
$\alum$                & $1.8\pm 4.2$ & $0.38\pm 0.50$ & $-1.9\pm 0.3$ & $0.0\pm 0.5$ \\
$ A_m^2\apol $         & --- & $-0.35\pm 0.01$ & $0.31\pm 0.13$ & $0.10\pm 0.13$ \\ \hline
$A_m$ Correction       & $0.0\pm 4.5$ & $1.0\pm 0.8$ & $2.75\pm 0.79$ & $0.76\pm 0.75$ \\ \hline
\end{tabular} \end{center}
\label{tab:alr-corr}
\end{table}\vskip -0.4in

\begin{table}[htbp]
\begin{center}
\parbox{6.0in}{\caption{Summary of SLD $\alr$ results.
Analysis of the 1996 run is still in progress.}\medskip}
\begin{tabular}{|l|c|c|c|c|l|} \hline
Run & Z Events & $\pola$ (\%) & $\sqrt{s}$ (GeV)& $\alr$ & Reference \\ \hline
1992 & $10.2\times 10^3$ & $22.4\pm 0.7$ & $91.55$ & $0.100\pm 0.044\pm 0.004$ 
& PRL 70, 2515\\
1993 & $49.4\times 10^3$ & $62.6\pm 1.1$ & $91.26$ & $0.1628\pm 0.0071\pm 0.0028$
& PRL 73, 25\\
1994-5& $93.6\times 10^3$ & $77.2\pm 0.5$ & $91.28$ & $0.1512\pm 0.0042\pm 0.0011$
& PRL 78, 2075\\
1996 & $ 51.4\times 10^3$ & $76.5\pm 0.8 $& $91.26$ & $0.1541\pm 0.0057\pm 0.0016$  
& preliminary \\ \hline
\end{tabular}
\end{center}
\label{tab:alr-summary}
\end{table}\vskip -0.2in

Each run has been carried out at slightly different center-of-mass energies.
To be comparable to other measurements and to extract $\swein$, the
$\alr$ measurement must be corrected to the Z-pole. Hence, combining the
above measurements and correcting to the Z-pole, including the small
$\gamma$-$Z$ interference correction, results in the cumulative result:
\begin{equation}
\alr^0 = 0.1550\pm 0.0034 \> \end{equation}
The corresponding value of $\swein$, assuming the SM relationship of 
Eq. \ref{eq:alr}, is $\swein = 0.23051\pm 0.00043$.
Finally, one can include the other less precise SLD lepton measurements
of $\swein$ due to the angular distributions of electron, muon, and tau-pair
events, as well as the so-called $Q_{LR}$ measurement\cite{Qlr-PRL}.
The combined result, still preliminary due to the 1996 results,
is \begin{equation} \swein = 0.23055\pm 0.00041\> . \end{equation}

\subsubsection{LEP Comparison and Global Fits}

\begin{figure}[tbh]
  \epsfysize=2.4in
  \begin{center}
  \epsfbox[70. 70. 530. 340.]{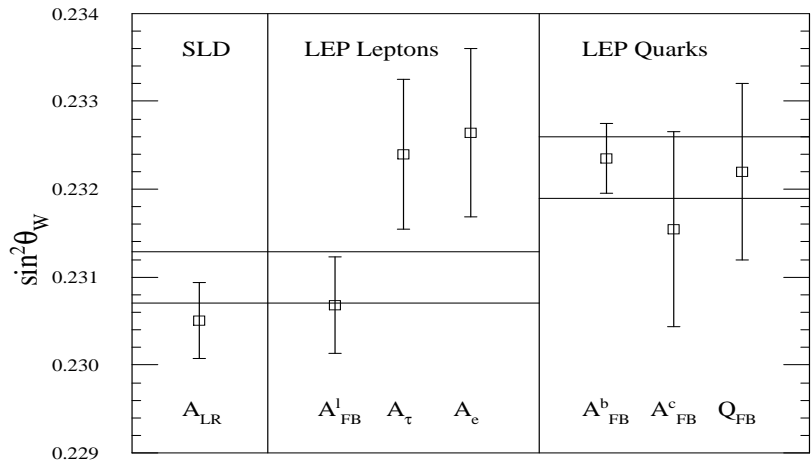}
  \parbox{5in}{\caption{Comparison of $Z^0$ measurements of $\swein$ for different
    techniques. The LEP results are from Moriond 1997.
    The horizontal bands represent the combined errors.}}
  \label{fig:swein-diffs}  \end{center}
\end{figure}

One way to compare the SLD $\alr$ measurement of $\swein$ with those
from LEP is shown in Fig. \ref{fig:swein-diffs}. The two most precise
measurements based on lepton couplings, $\alr$ and $A^\ell_{FB}$ (the
lepton forward-backward asymmetry from LEP), agree at the $0.2\sigma$
level. On the other hand the combined lepton-based measurements disagree
with the quark-based measurement (dominated by $A^b_{FB}$, 
the $b$-quark forward-backward asymmetry from LEP) by $2.7\sigma$.
It is interesting to speculate on whether the discrepancy represents a
somewhat unlikely, but surely plausible, distribution of values about a
central mean, or whether there is an experimental problem, or perhaps 
it is a hint of new physics. 

As stated earlier, $\alr$ is substantially a rather simple
event counting measurement. The one outstanding systematic is the scale of
the beam polarization. From Table \ref{tab:pol-sys} one finds that the
average error on the polarization scale is $1.0\%$, and this scale has
been independently cross-checked many times at the 1--3\% level.
On the other hand, to force the $\alr$ measurement into agreement with
the quark-based average of Fig.~\ref{fig:swein-diffs} would require a 9\%
uncertainty on the polarization scale. This is evidently not plausible.
Clearly, if the discrepancy 
between leptons and quarks were to become more significant, then the
implication would be for a non-SM explanation. 
 
We have performed some global fits of the electroweak data, involving the
fit of 5 parameters ($m_t$, $M_H$, $\alpha_s$, $\alpha(M_Z^2)$, $M_Z$)
to 19 observables. First of all, one can see how well the precision
measurements predict the top mass. The result is $155^{+11}_{-9}$ GeV/c$^2$,
which is about $1.7\sigma$ from the measured value. This is not excellent
agreement, but is probably acceptable. If the measured $m_t$ is incorporated
into the fit, one can derive a Higgs mass. The errors on the central value
are still huge, implying that it should not be taken too seriously.
However, the upper limit of $M_H<419$ GeV/c$^2$ at 95\% CL is
probably sufficiently well determined as to be meaningful.

\begin{figure}[bth]
  \epsfysize=2.5in
  \begin{center}
  \epsfbox[70. 40. 530. 340.]{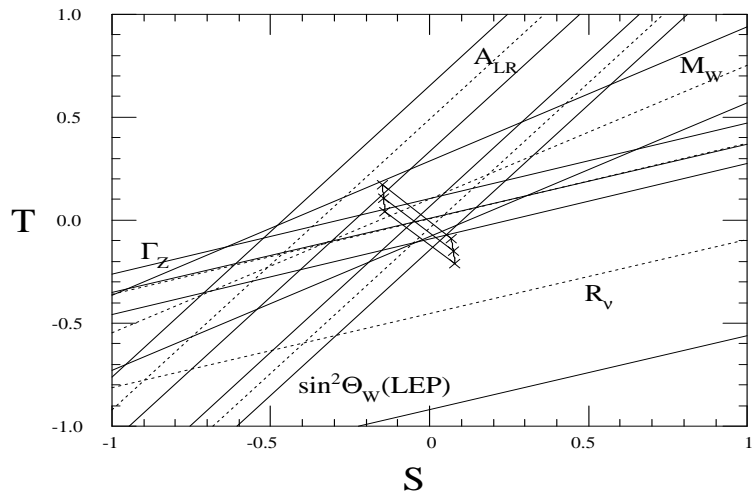}
  \parbox{5in}{\caption{Analysis of global results using the S-T framework. Central values
  for the measurement bands are the dotted lines. The SM region is given by
  the central region bounded by crosses (see text).}}
  \label{fig:ST}  \end{center}
\end{figure}

The $S$-$T$ framework\cite{ref:TakPesk}
provides a more general tool for SM tests.
Figure~\ref{fig:ST} indicates how several of the key measurements
translate into bands of different slope in the $S$-$T$ plane.
Also shown are the SM predictions (crosses) centered on the origin, with $T$
increasing with top mass ($175\pm 6$ GeV/c$^2$)
and $M_H$ variation along the diagonal lines (1000 GeV/c$^2$ lower right
to 70 GeV/c$^2$ upper left). There is
generally good agreement between the data and the SM, although it is
interesting to note that that there is a hint of a second solution for
the data at negative $S$ and $T$.

\section{The \zbb ~Vertex}

The process $Z^0\rightarrow b\bar{b}$ has a set of unique vertex corrections
involving the top quark and, potentially, physics beyond the SM, which are
quite different from the oblique corrections which modify the theoretical
value of $\swein$. For example, in addition to vertex correction diagrams
involving $t$-$b$-$W^\pm$ couplings, one can also have $t$-$b$-$H^\pm$. The primary
observables used to describe this physics are the fraction of hadronic $Z^0$
decays which are to $b\bar{b}$, $R_b=\Gamma_b/\Gamma_{\rm had}$, and the
$b$-quark parity violation parameter $A_b$. To lowest order, the SM expressions
for these quantities in terms of the axial-vector and vector couplings, or the
left and right-hand couplings, of the $b$-quark to the $Z^0$ are given by
\begin{equation}
R_b\propto v_b^2 + a_b^2 \propto g_L^2 + g_R^2\>;\hspace{0.3in} 
A_b = {{2a_ev_2}\over{a_e^2+v_e^2}} = {{g_L^2-g_R^2}\over{g_L^2-g_R^2}}
\end{equation}
With $g_L=-0.42$ and $g_R=0.08$ we see that $R_b$ and $A_b$ are quite
complimentary: $R_b$ is $5.5$ times more sensitive to $g_L$ than $g_R$, whereas
$A_b$ is $5.5$ times more sensitive to $g_R$ than $g_L$.

\subsubsection{Topological Vertexing and VXD3 Results}

\begin{figure}[tbh] \begin{center}
\makebox[2.9in]{
  \epsfxsize=2.8in 
  \epsfbox{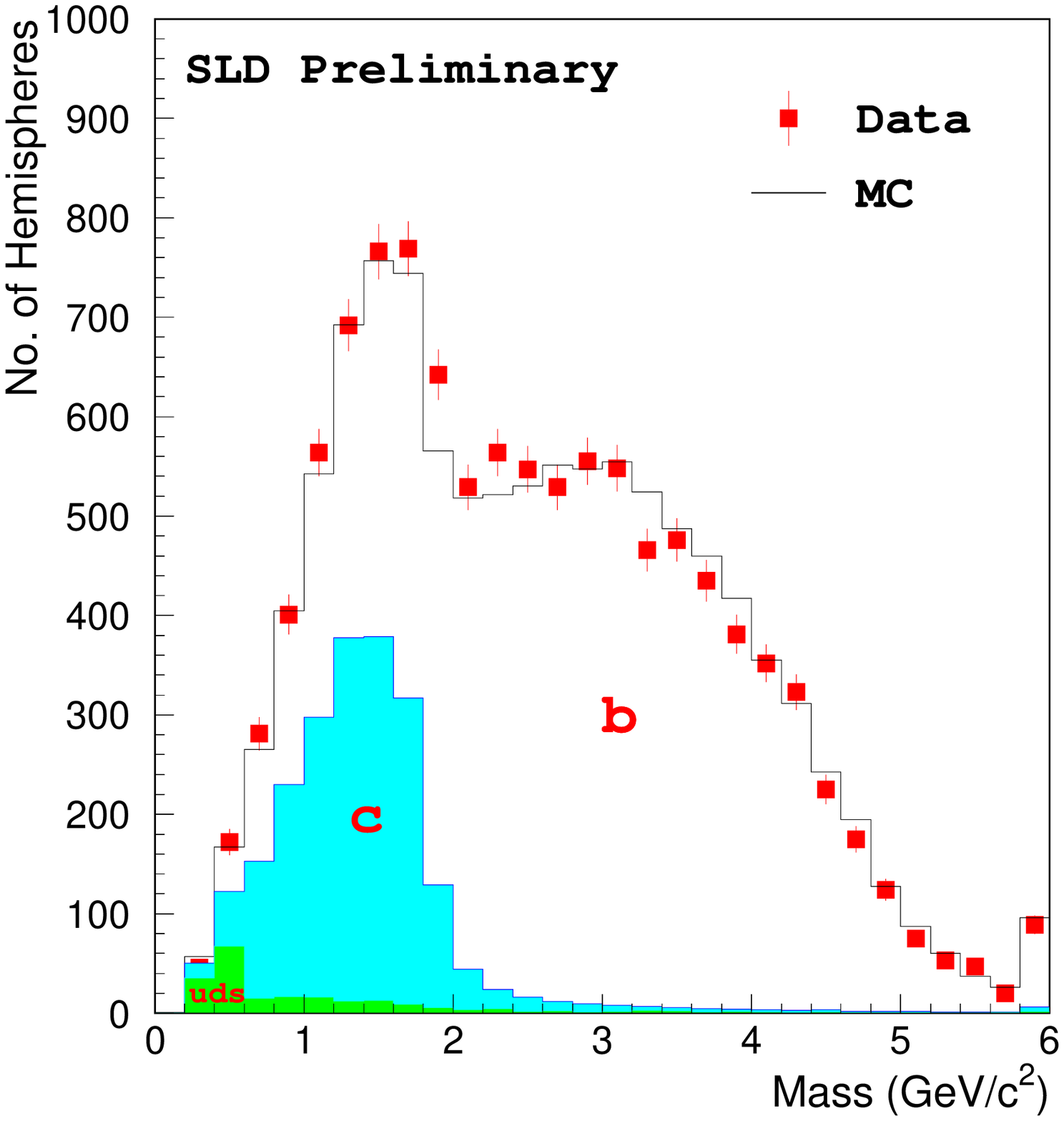}
 } \makebox[0.10in]{}
\makebox[2.9in]{
  \epsfxsize=3.0in
  \epsfbox{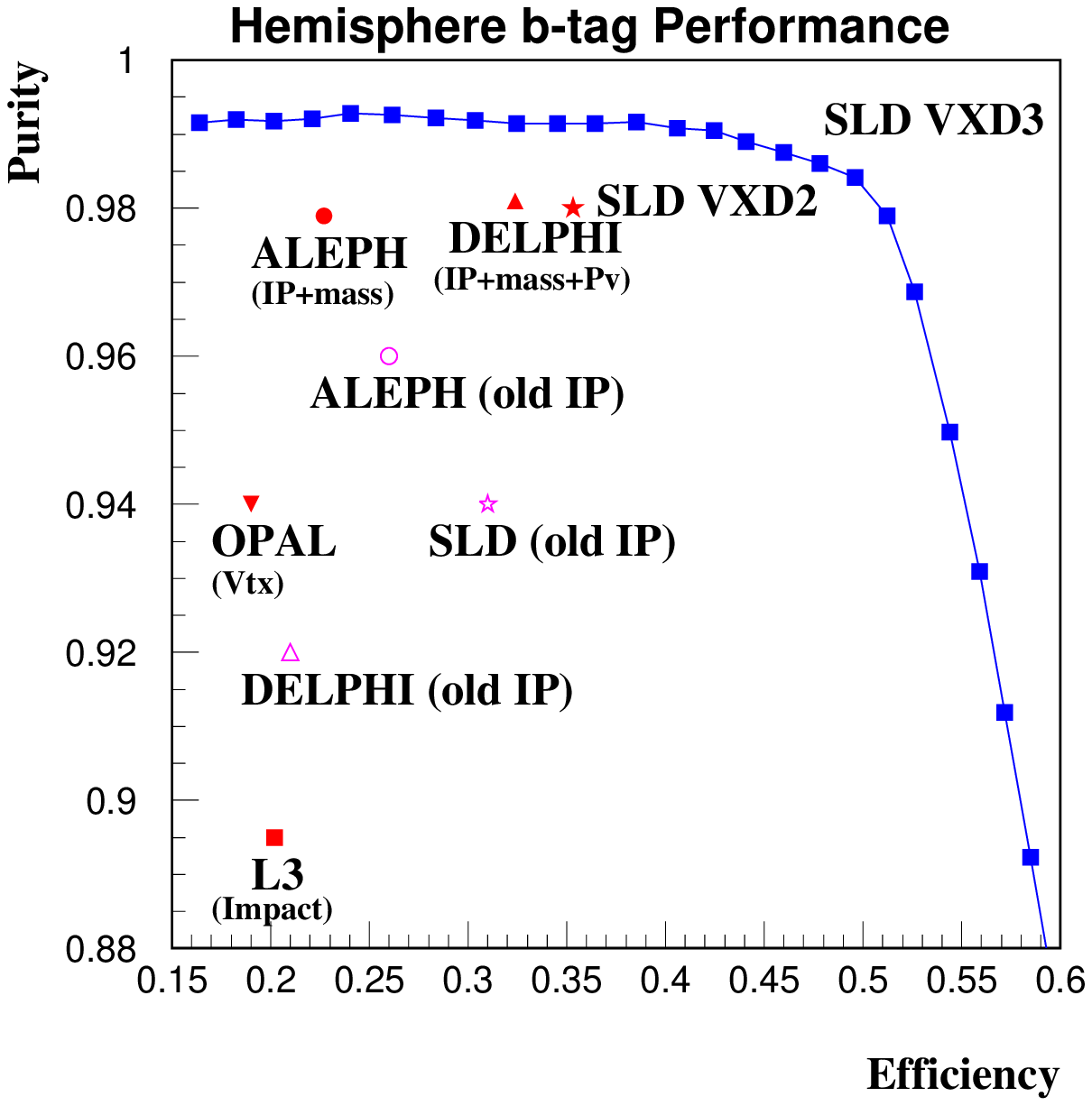}
 }
  \parbox{5.5in}{\caption{Left: Reconstructed mass of secondary vertex decay products using
   topological vertexing. The data points are for the 1996 run with VXD3.
   This agrees well with the MC, for which the expected contributions
   due to $b$, $c$, and $u,d,s$ are indicated. 
   Right: Hemisphere $b$-tag performance in 
   efficiency-purity plane for SLD. Recent improvements in tagging techniques
   from LEP (except for DELPHI) are not yet included in the plot. }}
  \label{fig:rb-mass} \end{center}
\end{figure}

The vertexing resolution available with SLD allows one to exploit
qualitatively new analysis techniques. The intrinsic resolution 
of VXD3 demonstrated with 1996 data is $4.7$ $\mu$m in $r$-$\phi$
and $4.5$ $\mu$m in $z$. Topological vertexing
refers to the reconstruction of displaced vertices and the tracks which
emanate from them. The tracks can then be combined and their properties
compared to Monte Carlo. In particular, the invariant mass of the tracks 
is a very powerful discriminant of bottom mesons relative to charm. In fact,
the kinematic information available can be used to partially compensate for
unmeasured (neutral) decay products. If $M_{\rm raw}$ is the mass from the
measured mass, the corrected mass is given by 
$M=\sqrt{M^2_{\rm raw} + p_T^2} + p_T$, where $p_T$ is the transverse
momentum of the visible tracks relative to the axis formed by the line from
primary collision point to reconsructed decay vertex. The resulting mass
distribution is given in Fig.~\ref{fig:rb-mass}, and is compared to Monte Carlo,
where one can see a clear separation between bottom and charm (as well as light
quarks). This particular distribution is for 1996 data with VXD3. This
separation is key for eliminating charm contamination and systematics
which have plagued $b$-quark measurements, in particular the $R_b$ measurement
described below.

\subsection{Direct $A_b$ and $A_c$ Measurements}
 
The differential cross section for $b$-quark production
with polarized beam is
\begin{equation}
{{d\sigma}\over{d\cos\theta}}\propto \left( 1-\pole A_e\right)\left(
1+\cos^2\theta\right) + 2\cos\theta\left( A_e - \pole\right)A_b
\label{eq:pol-afb}
\end{equation}
The beam polarization, $\pole$, in this expression can be of either sign.
There are two important aspects of the polarized angular asymmetries
which are readily apparent from this expression.
First, the $A_b$ term is modulated by the product $\pole A_b$. This is contrasted
with the unpolarized forward-backward asymmetry for which the
asymmetry term is modulated by the product $A_e A_b$. Hence, the SLD
measurement represents a direct $A_b$ measurement, rather than a product
of two SM parameters. Secondly, the much larger polarized asymmetry
implies a statistical advantage by a factor $(\pole/A_e)^2\approx 25$.

\begin{figure}[tbh] \begin{center}
\makebox[2.9in]{
  \epsfxsize=3.2in
  \epsfbox{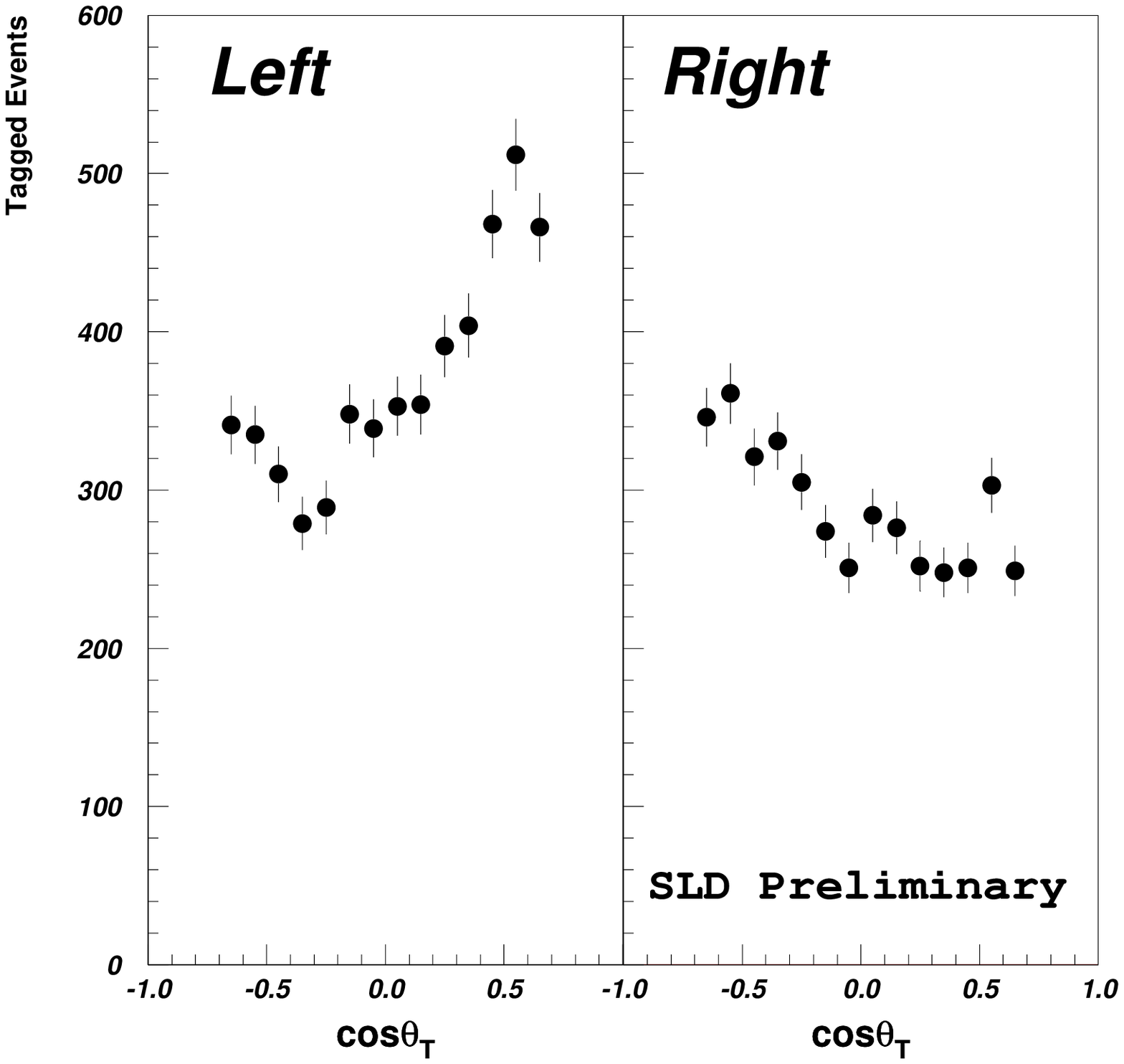}
 } 
\makebox[2.9in]{
  \epsfxsize=3.0in
  \epsfbox[30. 200. 560. 670.]{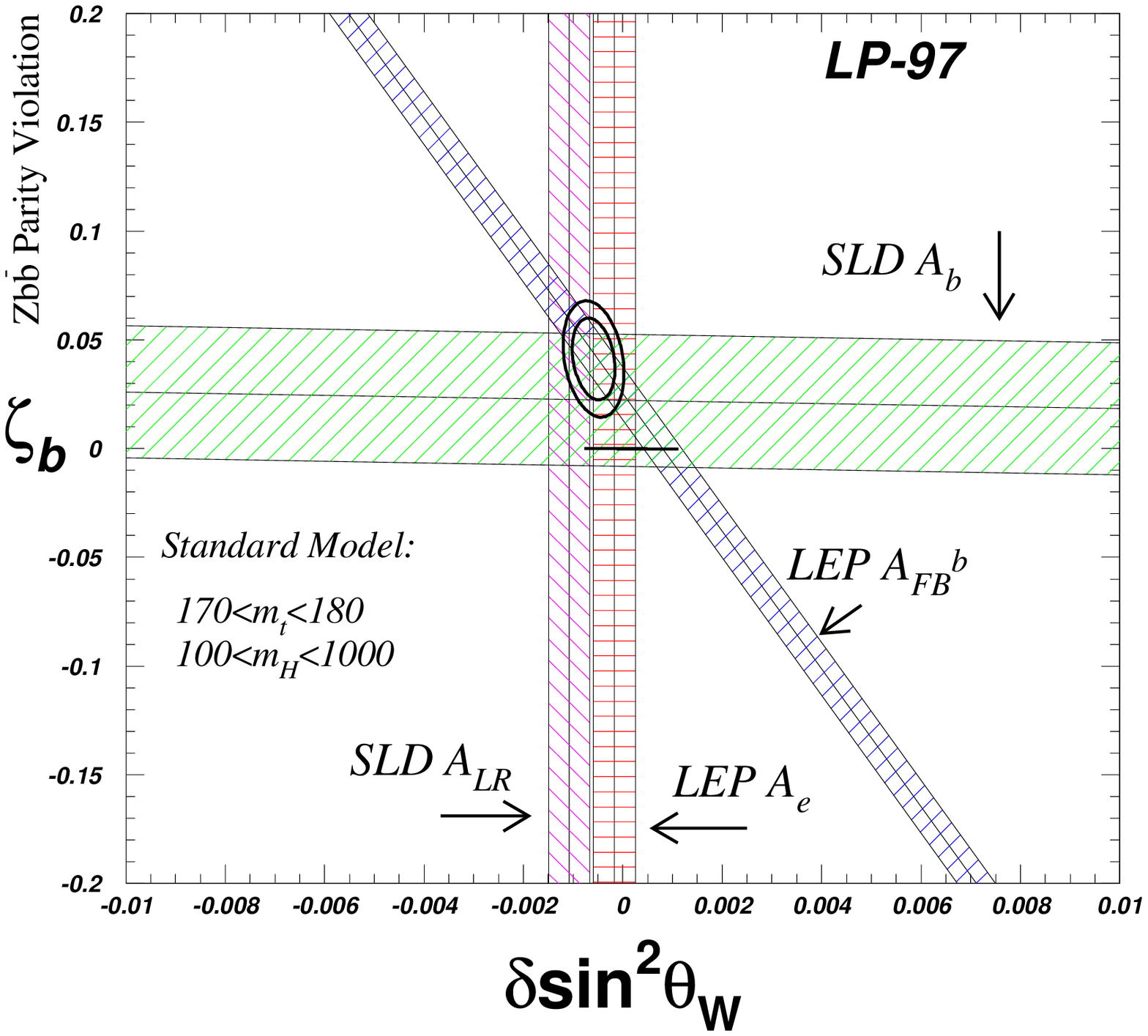}
 }\vskip 0.1in
  \parbox{5in}{\caption{Left: Angular distributions for $Z\rightarrow b\bar{b}$, plotted
   separately for left and right-hand polarized electron beam.
   Right: Analysis of Ref. 7 applied to the LEP-SLD measurements
   of $A_b$ and $\swein$. The horizontal line centered at the origin
   represents the SM prediction for the ranges $170<m_t<180$ GeV/c$^2$ and
   $100<M_H<1000$ GeV/c$^2$, and using as input $\alpha_s=0.117$ and
   $\alpha=1/128.96$.
   The ellipses are the combined experimental 68\% and 90\% CL error contours.}}
  \label{fig:Ab-data} \end{center}
\end{figure}

For this measurement vertexing is required only to identify $b\bar{b}$
final states, which is relatively straightforward. For previous
measurements this was accomplished using a minimum number of tracks
with a significant 2-D impact parameter relative to the primary collision
point. This is now done using the SLD topological vertexing
method, discussed in the preceding section.
Once the event is tagged, it is necessary to identify which jet is the
$b$-jet. This has been carried out in three analyses: using the jet charge
technique, using the sign of the charge of a high-$p_T$ lepton from
semi-leptonic $b$ or $\bar{b}$ decays, or using the sign of charge of
a $K^\pm$ due to $b\rightarrow c\rightarrow s$ and 
identified by the CRID system. The lepton\cite{ref:Ab-lepton} and kaon analyses
rely on Monte Carlo (MC) to determine the relative 
$u,d,s,c,$ and $b$ contributions
as well as the probability of correctly assigning $b$ and $\bar{b}$. On the
other hand the jet charge technique\cite{ref:Ab-jet} is self calibrating, with
very little dependence on MC. The left-hand plot of 
Fig.~\ref{fig:Ab-data} shows the distributions
from the 1994--95 run to which Eq. \ref{eq:pol-afb} is fit
to extract $A_b$. One clearly sees both
large left-right and forward-backward asymmetries. 
The $A_c$ measurment is performed analogously, where the final state
involves either sign of lepton charge in semi-leptonic charm decays 
or sign of charge of reconstructed $D$ or $D^*$ mesons.

\begin{figure}[tbh] \begin{center}
\makebox[2.9in]{
  \epsfxsize=2.9in
  \epsfbox[10. 100. 600. 770.]{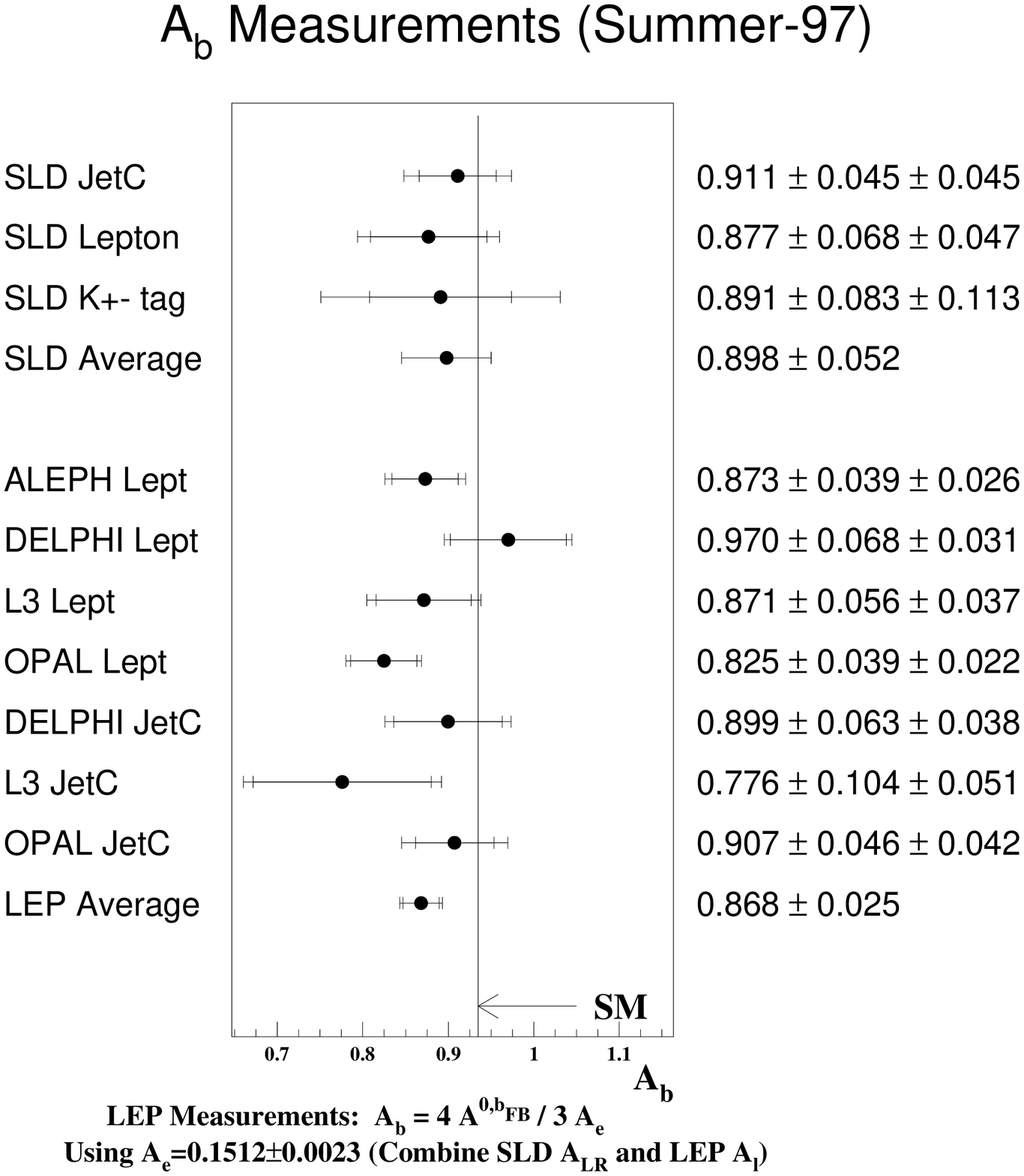}
 }
\makebox[2.9in]{
  \epsfxsize=2.9in
  \epsfbox[10. 100. 600. 770.]{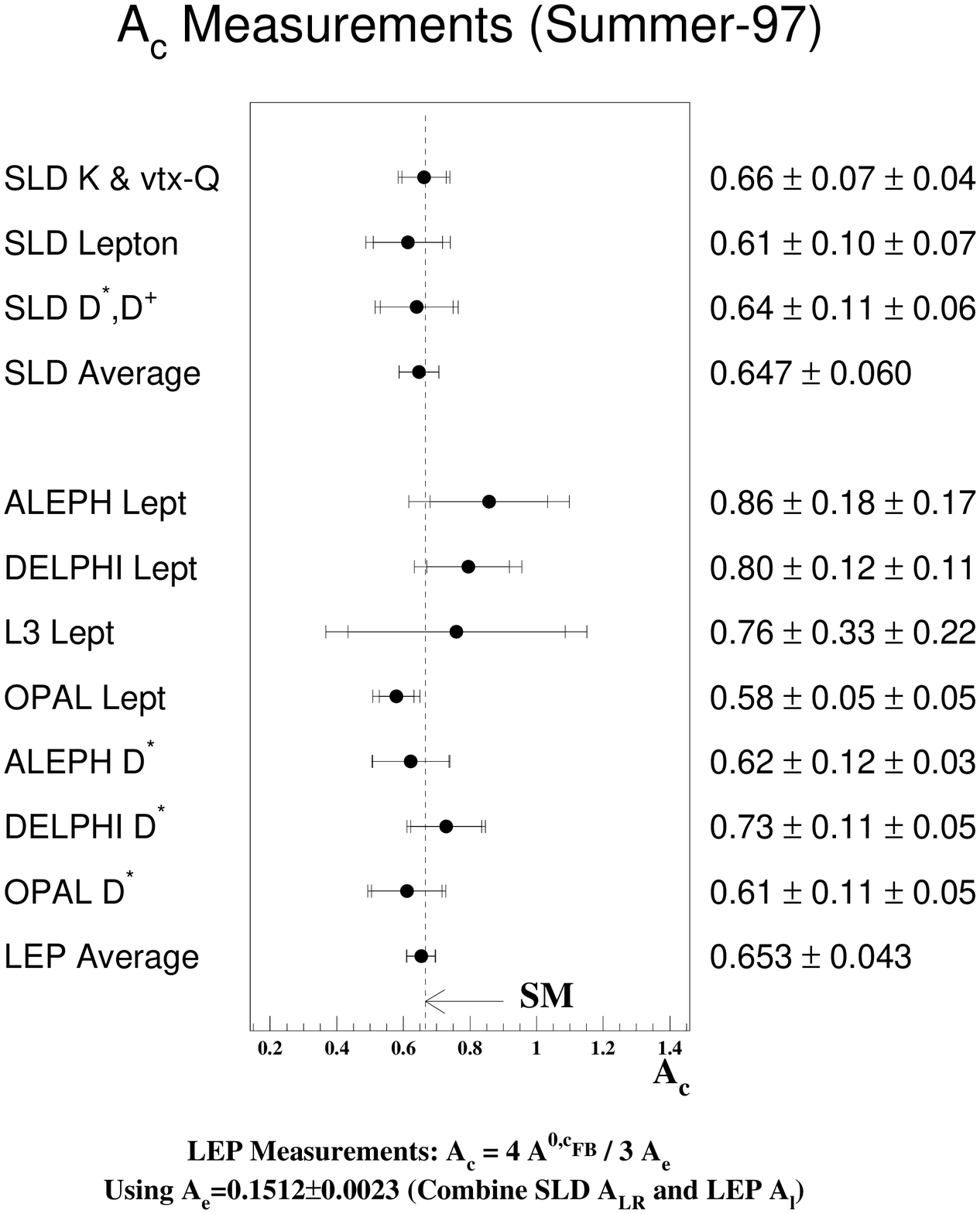}
 } \vskip 0.1in
  \parbox{5in}{\caption{Comparision of world $A_b$ (left) and 
   $A_c$ (right) measurements.}}
  \label{fig:ab-world} \end{center}
\end{figure}

A compilation of SLD and LEP results for $A_b$ and $A_c$ is given
in Fig.~\ref{fig:ab-world}.
The SLD kaon analysis has been updated since Winter 1996. As discussed above,
since the LEP experiments measure the product $A_e A_b$ ($A_e A_c$), 
a value of $A_e$ must be specified to extract $A_b$ ($A_c$). For these plots
the combined LEP-SLD value of $A_e = 0.1512\pm 0.0023$ has been used. 
One sees that for $A_b$ there is a substantial disagreement 
 of $2.7\sigma$ between the
experimental average and the SM prediction. 
It is sometimes convenient to express deviations from the SM \zbb ~couplings
as a linear combination of left and right-handed terms. Specifically,
the couplings $\xi_b$ and $\zeta_b$ are chosen\cite{ref:tgr} such that
$R_b$ depends only on $\xi_b$ and $A_b$ only on $\zeta_b$, and as such
$\zeta_b$ is independent of $m_t$. The right-hand plot of
Fig.~\ref{fig:Ab-data} gives $\zeta_b$
{\it vs} $\swein$ from LEP and SLD compared to the SM predictions, 
centered at the origin. The agreement of the SM to the combined
experimental result is rather poor.

\subsection{Measurement of $R_b$ and $R_c$}   

A serious challenge to the $R_b$ measurement has been the need to
accurately determine the charm background. In SLD, topological vertexing
combined with the mass selection of Fig.~\ref{fig:rb-mass} has greatly reduced
the charm contamination and hence obviated the reliance on troublesome MC
charm modelling. SLD applies this $b$-tag separately to each event hemisphere,
therefore measuring the rate of single tags and double tags. In this way, one
determines directly from the data not only the $b$ fraction, but also the
$b$-tag efficiency. The right-hand plot of
Fig.~\ref{fig:rb-mass} shows the SLD (VXD3) 
contour for hemisphere $b$-tag measurement in the efficiency-purity plane.

SLD also uses the vertex double-tag technique for its $R_c$ measurement,
and is, so far, unique in doing so. The two major $c$-tag selection criteria
are a vertex mass cut ($0.6<M<2.0$ GeV/c$^2$) within the charm region of 
Fig.~\ref{fig:rb-mass} and a cut on the momentum of the reconstructed decay
vertex, which exploits the fact that for a given vertex mass, $D$ mesons
have a larger momentum than do $B$ mesons. The resulting hemisphere $c$-tag
efficiency and purity (with VXD3) are $14.0\pm 1.3\%$ and $67\pm2\%$,
respectively.

\begin{figure}[tbh] \begin{center}
\makebox[2.9in]{
  \epsfxsize=2.9in
  \epsfbox[10. 100. 600. 770.]{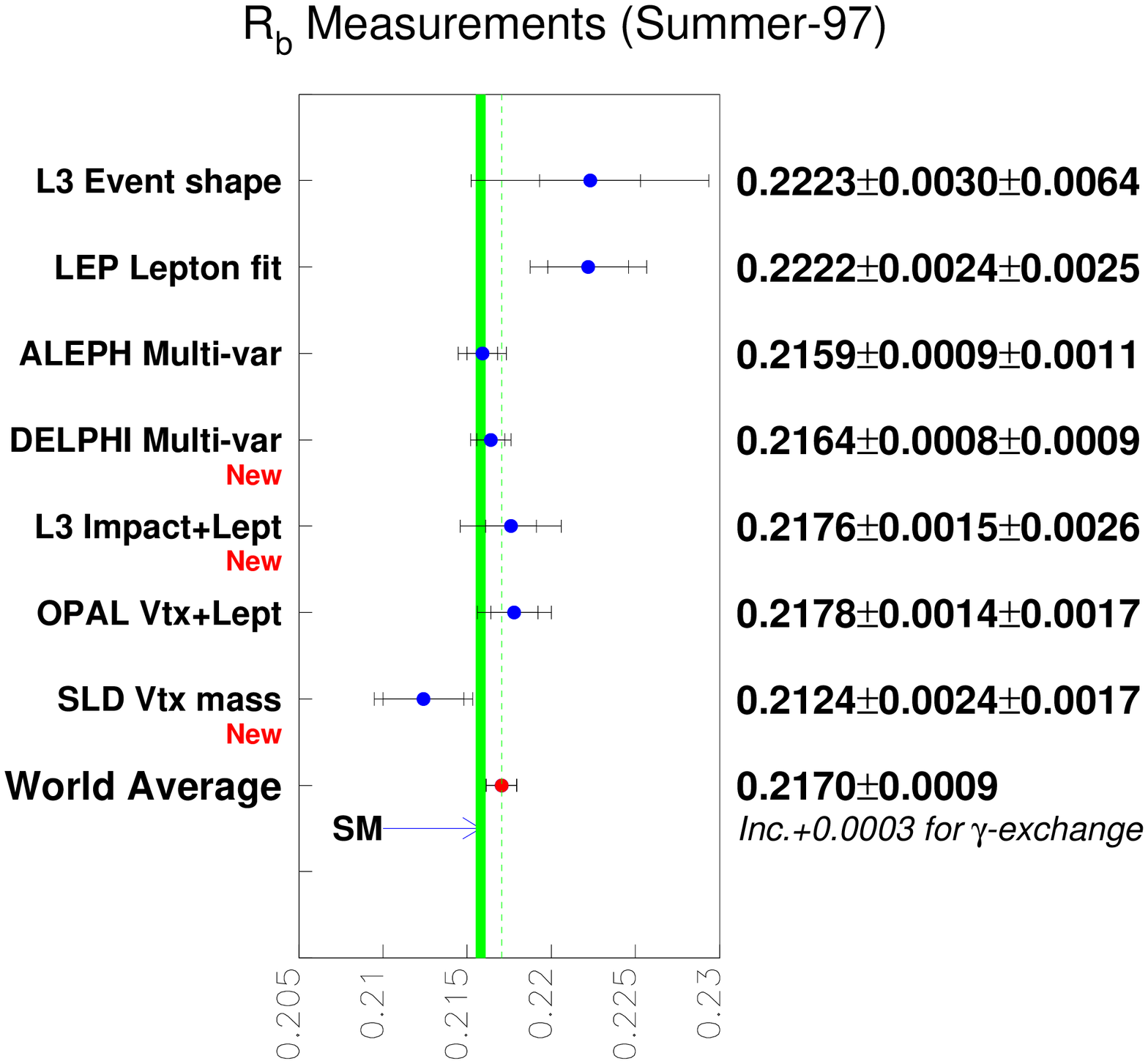}
 }
\makebox[2.9in]{
  \epsfxsize=2.9in
  \epsfbox[10. 100. 600. 770.]{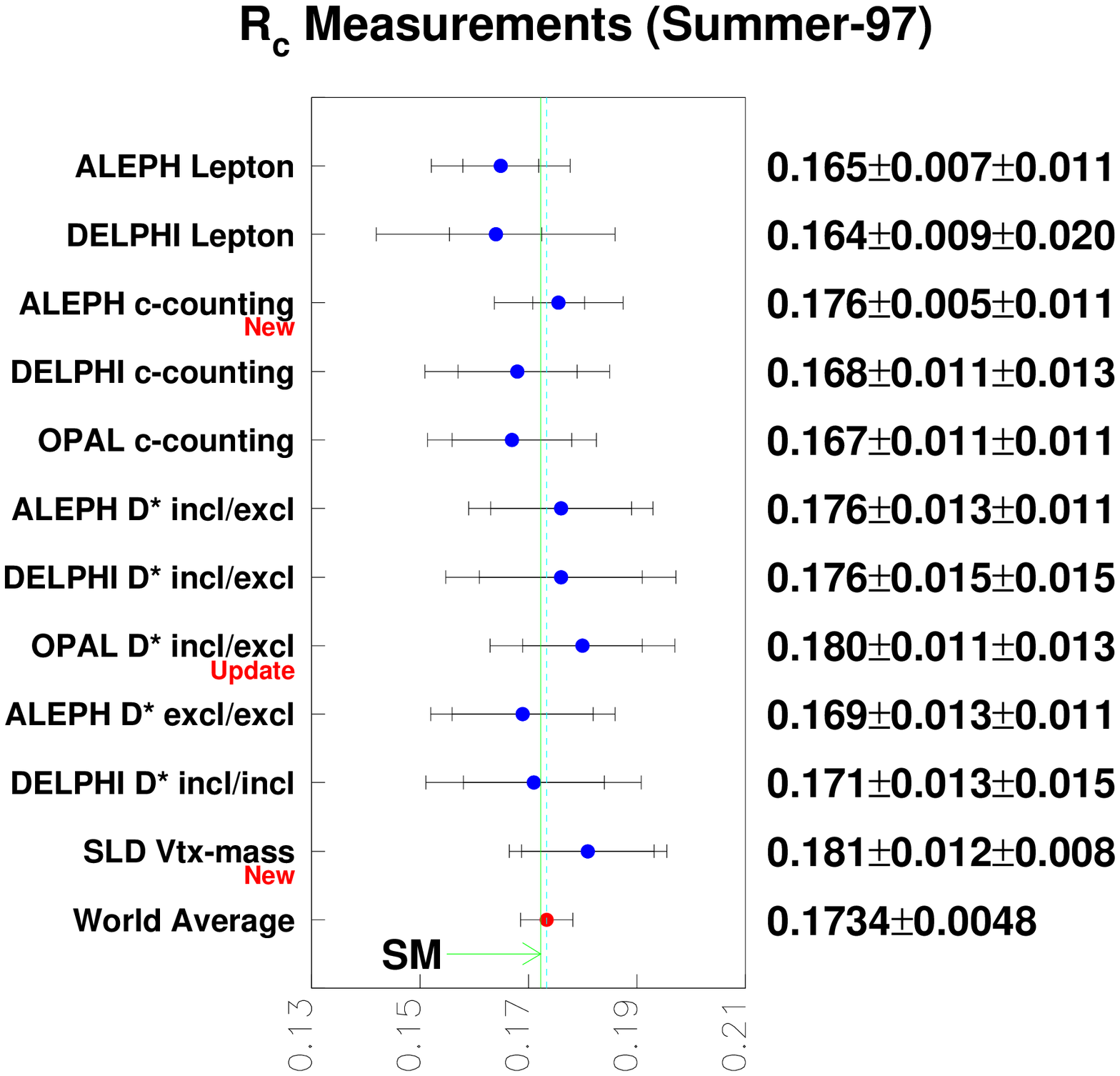}
 } \vskip 0.05in
  \caption{Comparision of world $R_b$ (left) and $R_c$ (right) measurements.} 
  \label{fig:rb-world} \end{center}
\end{figure}

A compilation of SLD and LEP results for $R_b$ and $R_c$ is given
in Fig.~\ref{fig:rb-world}. The measurements from both SLD and LEP have
improved substantially since 1995, when there was an apparent $3\sigma$
discrepancy in $R_b$ between experiment and the SM.  Reduction of
charm background using vertex mass, as done by SLD and ALEPH, along
with better treatment of hemisphere correlations by all groups, has
improved the experimental result. And now the SM prediction lies within
the 68\% CL experimental error ellipse.

\section{$B$ Mixing}

Measurement of $B^0$-$\bar{B^0}$ mixing provides an excellent means for
testing the SM, especially the CKM sector and
the SM description of CP violation. SLD and all LEP experiments have 
reported measurements of $B_d^0$ mixing, as outlined below.
No measurement has yet been made of $B_s^0$ mixing, although there are
limits from LEP. Because of the rapid oscillations expected in the $B_s$ case,
vertexing resolution is very important. In addition to its excellent vertexing,
SLD also benefits greatly from polarized beam for initial-state tagging. Hence,
if SLC can deliver good luminosity, then SLD will be well
positioned to make this difficult measurement.

The considerable interest in making a $B_s^0$ mixing measurement
derives largely from the following considerations.
The mixing box diagrams give the following expression for the measureable
ratio of mass eigenstate splitting for $B_s$ relative to $B_d$:
\begin{equation}
{{\Delta m_s}\over{\Delta m_d}} = {{\eta_s M_s f_s^2 B_s}\over{
\eta_d M_d f_d^2 B_d}}{|{V_{ts}|^2}\over{|V_{td}}|^2} = 
(1.32\pm 0.09){M_s\over M_d}{|{V_{ts}|^2}\over{|V_{td}}|^2}
\end{equation}
Here, $\eta$, $M$, $f$, and $B$ represent, respectively, the 
QCD correction, the mass, 
the decay constant, and the degree to which the mixing occurs via box
diagrams. 
The quantities $f$ and $B$ in particular are not well determined. However in ratio
they are determined to $< 10\%$ from lattice calculations, as
indicated by the numerical value given\cite{ref:bs-lattice}. Since $V_{ts}$
is derived from unitarity, the $B_s$ mixing measurement will present a
good measure of what is the most poorly known side of the unitarity triangle,
$V_{td}$. Hence, the measurement will pinpoint the apex 
$(\rho , \eta)$ of the unitarity
triangle and, in fact, provide a substantial test of the SM picture of CP
violation by checking that, in fact, such a triangle exists.

\subsection{$B_d$ Mixing Measurement}

\begin{figure}[tbh] \begin{center}
  \makebox[2.9in]{
  \epsfxsize=3.1in
  \epsfbox{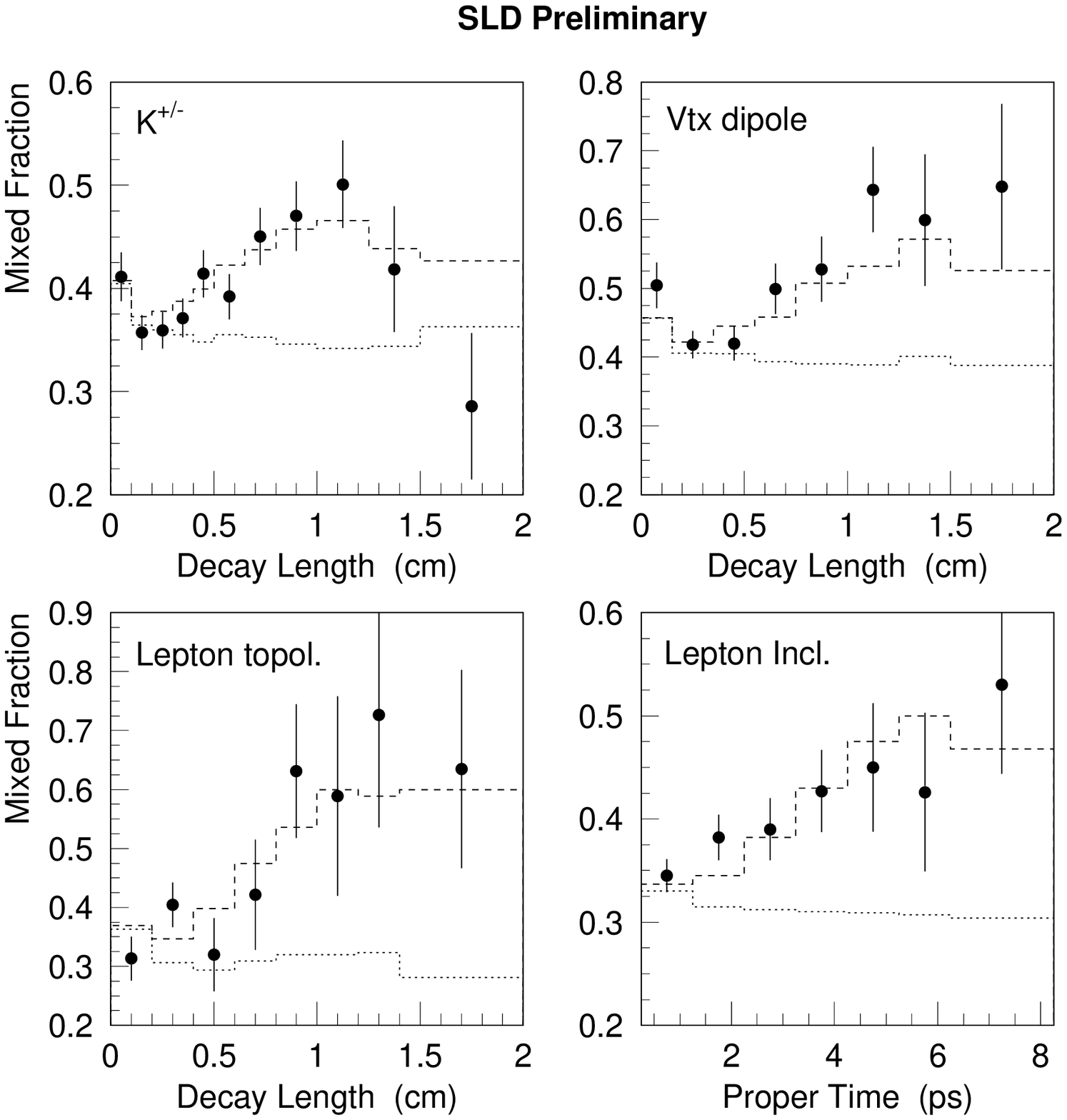}
  } \makebox[2.9in][r]{
  \epsfxsize=2.7in
  \epsfbox{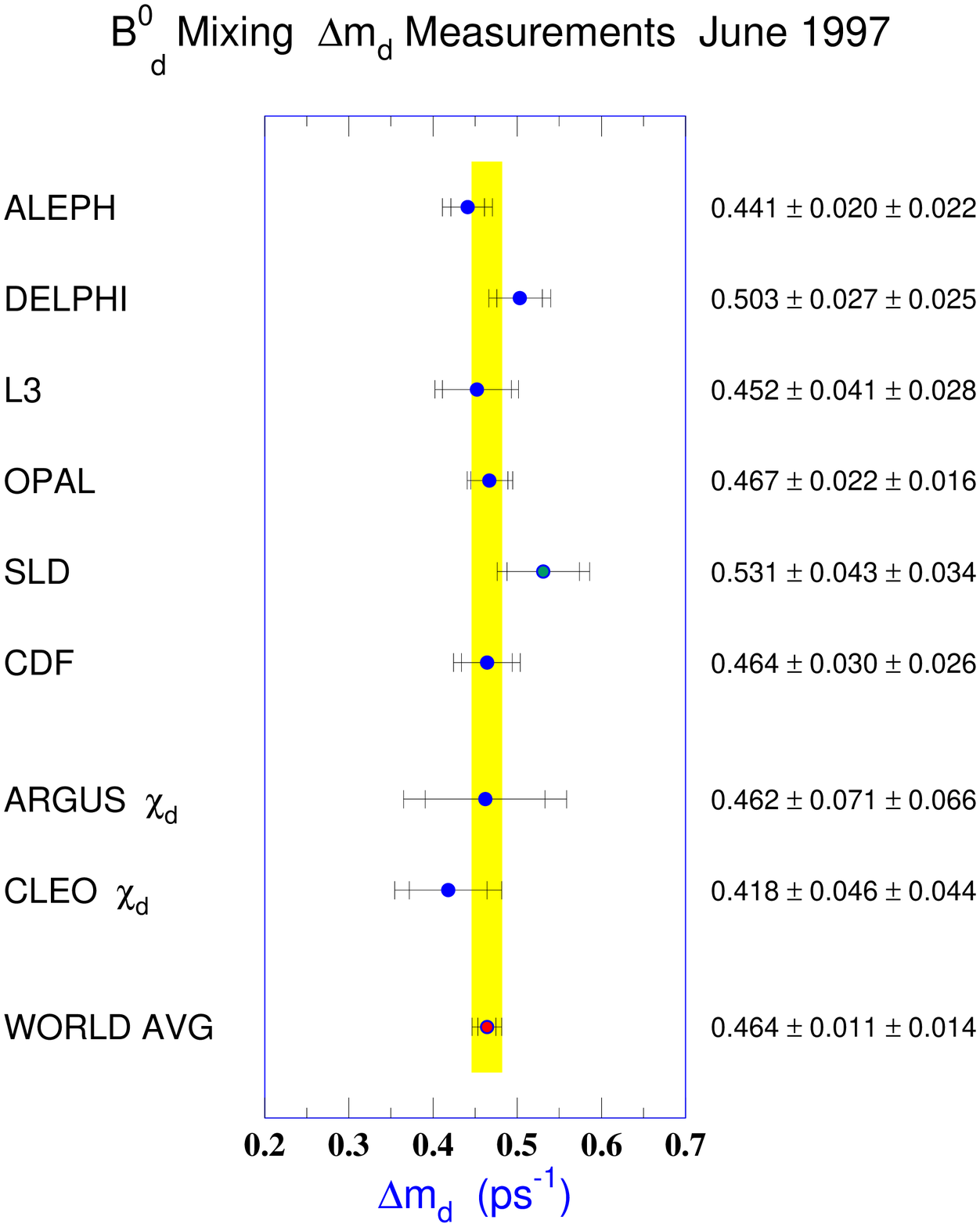}
  } \vskip 0.12in
  \parbox{5in}{\caption{Left: SLD results for $B_d^0$ mixing for the four different
   analysis techniques used for final-state tagging. The dotted lines
   correspond to the MC result with no mixing, while the dashed line
   is the fitted mixing result. Right: Comparison of world $B_d^0$
   mixing results.}} 
  \label{fig:bd4} \end{center}
\end{figure}

Once a sample of $Z\rightarrow b\bar{b}$ events has been identified,
a mixing measurement involves three major requirements:
\begin{enumerate}
\item The proper time between the $B^0$ production at the collision point
and the $B^0$ (or $\bar{B^0}$ if mixing has occurred) decay vertex. This is
determined from the observed decay length and the measured boost.
\item An initial-state tag to determine $B^0$ {\it vs} $\bar{B^0}$ at production.
\item A final-state tag to determine if mixing has occurred by the
  time of decay.
\end{enumerate}
The large forward-backward asymmetry for $Z\rightarrow b\bar{b}$ with
polarized beam is measured to extract $A_b$, as described above. However,
this asymmetry can now be exploited to tag the initial-state flavor of a
$b$ jet from its production angle. One can combine this with the jet-charge
technique to produce an initial-state tag which is correct with a
probability of $\approx 74\%$ for $\pole = 77\%$.
The final state tag is carried out using four different techniques in SLD.
Two of these involve using semi-leptonic decay events, one using
a reconstructed $D$ meson and the other being more inclusive. The third uses
the charge of a $K^\pm$ as identified by the CRID system. This is similar
to that used for the $A_b$ analysis. Finally, there is a 
``charge dipole'' tag, which
uses 3-D topological vertexing to look for a charge separation (dipole)
due to the $b\rightarrow c$ decays. The mixing signals for all four of these
techniques are shown separately in Fig.~\ref{fig:bd4}.

\subsubsection{$B_s$ Mixing Prospects}

The SLD analysis for $B_s^0$ mixing is similar to that described above
for $B_d^0$. Initial-state tagging is the same, and the final-state tag
techniques are similar.
The primary qualitative difference is the rapid oscillations,
where the mixing parameter 
$x_s = \Delta m_s/\Gamma_s = 1.55\,\Delta m_s $ is already 
known to be greater than about 14, implying
many mixing oscillations within a decay lifetime. This
puts a premium on vertex detector resolution. 
In particular, the ``charge dipole'' final-state tagging
technique mentioned above actually doubles in statistical power
relative to the $B_d^0$ case.

The estimate for the SLD reach with 
$\sim 500\times 10^3$ $Z^0$ events (with VXD3),
when expressed as a limit, is $\Delta m_s < 18$ ps$^{-1}$ at 95\% CL.
This is significant, as it covers essentially the entire presently allowed
region for the apex point $(\rho ,\eta)$ of the unitarity triangle. 
Hence, given enough luminosity, SLD would be guaranteed to see the 
mixing or place very interesting limits on it.

\section{Summary and Prospects}

The SLD $\alr$ measurement continues to be the single most precise
measurement of $\swein$, with the cumulative result being
$\swein = 0.23051\pm 0.00043$. This agrees well with the cumulative
LEP result derived from lepton couplings, but disagrees with the
LEP quark-derived value of $\swein$ at $3.1\sigma$, and differs from
the global LEP average by $2.8\sigma$.
The polarized forward-backward asymmetries allow direct measurement
of $A_b$ and $A_c$, which are mostly sensitive to the right-hand vertex
couplings. The world data for $A_b$ presently disagree with the SM at the
$2.7\sigma$ level. The SLD measurement of
$R_b$, after careful elimination of charm with precision vertex
measurement and topological vertex reconstruction, is in good
agreement with the SM. 

The error on the measurement of $\alr$ is dominated by the event
statistics, $N_Z$, and this will continue into the future. 
To excellent approximation the error on $\swein$ from $\alr$ is
\begin{equation}
\delta\swein = {1\over 7.9}\left[ {1\over{N_Z \pole^2}} + 
A_{LR}^2\Bigl({\delta\pole\over\pole}\Bigr)^2\right] ^{1/2}\> , 
\end{equation}\label{eq:alr-err}
where $\pole$ is the electron-beam polarization measurement, which is
the dominant systematic error.
Hence, one can extrapolate the error with much confidence. 
The other key measurements can also be extrapolated for
more data. This is summarized below in Table~\ref{tab:future}. The
most difficult to extrapolate is that for $B_s^0$ mixing.
Since some of these measurements, especially $\alr$ and
$A_b$ are currently in marginal disagreement
with the Standard Model, it is certainly of great interest to 
improve the errors and determine if there really is, or is not, an
indication of physics beyond the Standard Model. In addition, SLD has 
unique sensitivity to $B_s$ mixing, for which the measurement would
make a substantial test of the SM model of CP violation.

\begin{table}[tbh] \begin{center}
\parbox{5in}{\caption{Current errors for a few key electroweak observables for 
LEP and SLD now, and estimates for SLD near future (1998)
and assuming additional running.}\smallskip}
\begin{tabular}{|c|c|c|c|c|} \hline
Quantity & LEP  1996 & SLD 1996 & SLD 1997-98 & SLD+ \\
         & (combined) & & ($\sim150$k Zs) & 350k more Zs \\
\hline\hline
$\sin^2\theta_W^{eff}$ & $\pm 0.00027$ & $\pm 0.00041$ & $\pm 0.00029$ & $\pm 0.00021$ \\
\hline
$ A_b $ & $\pm 0.025$ & $\pm 0.040$ & $\pm 0.032$ & $\pm 0.023$ \\
\hline
$ R_b $ & $\pm 0.0012$ & $\pm 0.0027$ & $\pm 0.0020$ & $\pm 0.0014$ \\
\hline
$ \Delta m_s$ (ps$^{-1}$) & 8--10 & in progress & 12 & 18 \\
(95\% CL) & & & & \\ \hline
\end{tabular} \end{center}
\label{tab:future}
\end{table} 
 

\vskip -0.2in

\begin{thebibliography}{99}
\bibitem{ref:pol} M. Woods, AIP Conference Proc. 343, 230 (1995).
\bibitem{ALR-PRL93} SLD Collab., Phys. Rev. Lett. 70, 2515 (1993).
\bibitem{ALR-PRL94} SLD Collab., Phys. Rev. Lett. 73, 25 (1994).
\bibitem{ALR-PRL97} SLD Collab., Phys. Rev. Lett. 78, 2075 (1997).
\bibitem{Qlr-PRL} SLD Collab., Phys. Rev. Lett. 78, 17 (1997).
\bibitem{ref:TakPesk} T. Takeuchi and M. Peskin, Phys. Rev. D46, 381 (1992).
\bibitem{ref:tgr} T. Takeuchi, A.K. Grant, and J.L. Rosner, Proc. DPF94, 1231 (1994),
    \\ \hspace{0.3in} hep-ph/9409211.
\bibitem{ref:Ab-lepton} SLD Collab., Phys. Rev. Lett. 74, 2895 (1995).
\bibitem{ref:Ab-jet} SLD Collab., Phys. Rev. Lett. 74, 2890 (1995).
\bibitem{ref:bs-lattice} A. Abada, {\it et al.,} Nucl. Phys. B376, 172 (1992); \\
      \hspace{0.3in} A. Soni, Brookhaven preprint BNL-61378 (1995). 
\end{thebibliography}
\end{document}